\documentclass[useAMS,usenatbib]{mnras}
\usepackage{newtxtext,newtxmath}
\usepackage[T1]{fontenc}
\DeclareRobustCommand{\VAN}[3]{#2}
\let\VANthebibliography\thebibliography
\def\thebibliography{\DeclareRobustCommand{\VAN}[3]{##3}\VANthebibliography}

\usepackage{orcidlink}
\usepackage{graphicx}
\usepackage{subfig}
\usepackage{subcaption}
\usepackage{amsmath}	
\usepackage{fancyhdr}
\usepackage{hyperref}
\usepackage{todonotes}
\usepackage{academicons}
\usepackage{orcidlink}
\usepackage{xcolor}
\usepackage{soul}
\sethlcolor{yellow}

\title[Phase Space of Low mass compact objects]{The Phase Space of  Low-Mass Binary Compact Objects from LIGO-Virgo-KAGRA Catalog: Hints on the Chances of Different Formation Scenarios}

\author[Afroz and Mukherjee]{
  Samsuzzaman Afroz $^{1}$\orcidlink{0009-0004-4459-2981} \thanks{samsuzzaman.afroz@tifr.res.in},
  Suvodip Mukherjee $^{1}$\orcidlink{0000-0002-3373-5236} \thanks{suvodip@tifr.res.in} 
  \\
  $^{1}$Department of Astronomy and Astrophysics, Tata Institute of Fundamental Research, Mumbai 400005, India
}

\date{Accepted XXX. Received YYY; in original form ZZZ}
\pubyear{2025}

\begin{document}
\label{firstpage}
\pagerange{\pageref{firstpage}--\pageref{lastpage}}
\maketitle

\begin{abstract}
 Gravitational wave (GW) observations have significantly advanced our understanding of binary compact object (BCO) formation, yet directly linking these observations to specific formation scenarios remains challenging. The BCO phase space provides a robust and data-driven approach to discover the likely formation scenarios of these binaries. In this study, we expand the previously introduced binary black hole phase space technique to encompass low-mass compact objects (LMCOs), establishing a novel framework to investigate their diverse formation mechanisms. Applying this approach to selected low-mass events $(\lesssim 5 M_\odot)$ from the GWTC-3 catalog and the recently observed GW230529 event, we show for the first time the phase space demonstration of the LMCOs and find the associated probabilities for different formation scenarios including neutron star, astrophysical black hole, or primordial black hole. Our analysis includes the astrophysical modelling uncertainties in and how it causes degeneracy between different formation scenarios. In future, with improvements in GW detector sensitivity and with detection of more GW events, the LMCO phase space framework will significantly strengthen our capacity to associate more likely formation scenarios over the other, thereby refining our understanding of compact object formation for both astrophysical and primordial scenarios, and its evolution across the cosmic redshift.
\end{abstract}


\section{Introduction}
The observation of gravitational wave (GW) events has profoundly advanced our understanding of the universe by providing unprecedented insights into the population of binary compact objects (BCOs) in cosmological distances \citep{Bailes:2021tot, Arimoto:2021cwc}. These systems carry distinct signatures reflective of their formation channels and evolutionary histories. Careful analysis and interpretation of these signals are essential for deepening our understanding of compact binary evolution \citep{Dominik:2012kk, LIGOScientific:2016aoc, Barrett:2017fcw, LIGOScientific:2018mvr, Mapelli:2021taw,Bouffanais:2021wcr,Tiwari:2020otp,Tiwari:2021yvr, Iorio:2022sgz,Franciolini:2021tla, KAGRA:2021duu,Cheng:2023ddt,Antonelli:2023gpu,2023ApJ...950..181W}.

In our recent work \citep{Afroz:2024fzp}, we proposed the \texttt{BCO Phase Space}, an innovative framework designed to characterize different formation scenarios of BCOs by mapping theoretical evolutionary trajectories directly onto the space defined by observable GW parameters, such as component masses, spins, and luminosity distances. By examining the overlaps between these theoretical trajectories and observational regions occupied by GW events, the \texttt{BCO Phase Space} method provides a powerful tool for identifying plausible formation channels. Our first study demonstrated this method on binary black hole (BBH) systems and obtained for the first-time probability of their association with different astrophysical and non-astrophysical formation channel scenarios. The dimension of the phase space can be expanded by including other astrophysical properties such as eccentricity, recoil velocity to identify different phase space distribution of BCOs.  

\begin{figure*}
\centering
\includegraphics[width=14cm, height=5.3cm]{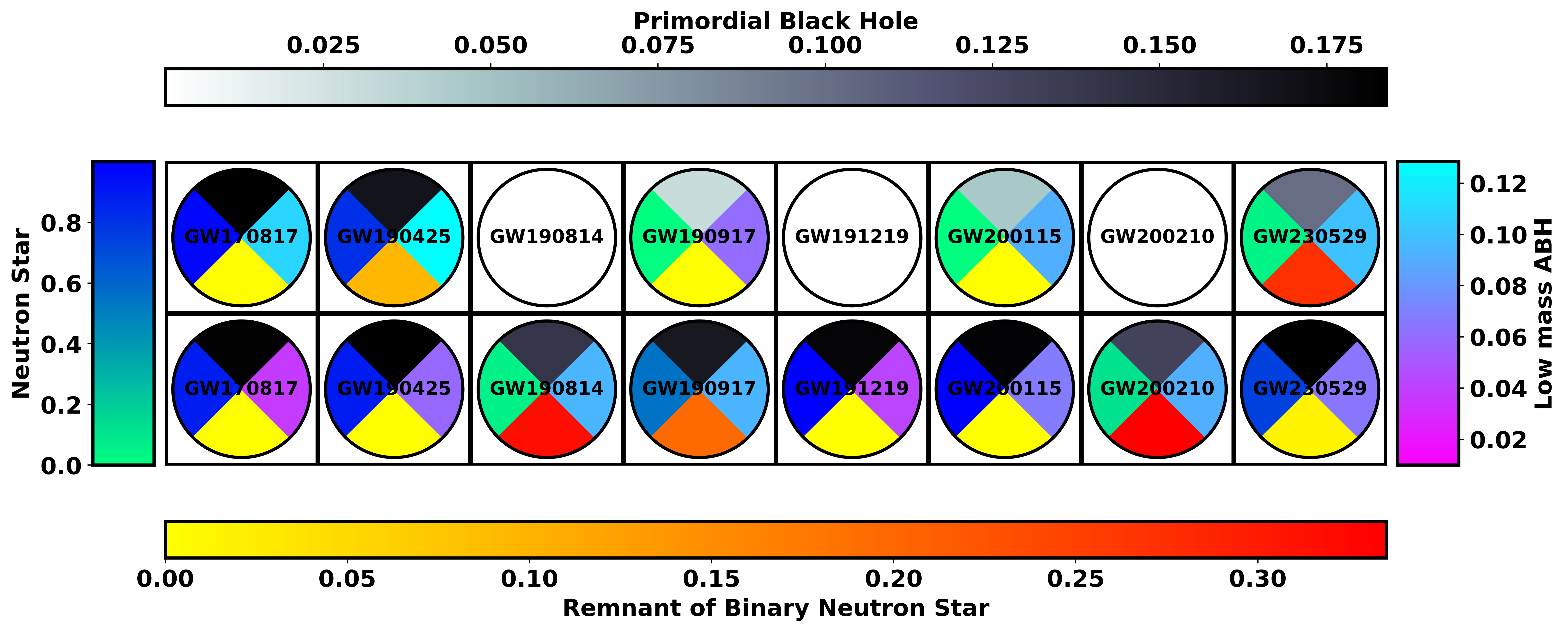}
\caption{The probabilities of seven low-mass GW events from GWTC-3 along with GW230529, evaluated based on their mass and spin posteriors. Each GW event is represented by two rows of pie charts corresponding to the primary (top row) and secondary (bottom row) components. Each pie chart is divided into four segments indicating the likelihood of different astrophysical formation channels: Neutron Star (left wedge, vertical left colorbar), Low mass Astrophysical Black Hole (right wedge, vertical right colorbar), Primordial Black Hole (top wedge, horizontal top colorbar), and Remnant of Binary Neutron Star (bottom wedge, horizontal bottom colorbar). Colors represent the relative probability weights assigned by integrating over the joint posterior distributions for mass and spin parameters against theoretical distributions for each astrophysical channel. These values are not expected to sum to unity for any individual event, as they are not normalized per event. They are normalised in the entire phase-space volume. As a result, the event-level probabilities should be interpreted within that channel-specific context. Three primary wedges are rendered in white as the object falls outside our low-mass compact-object criteria.}
\label{fig:ProbEvents}
\end{figure*}

In this current study, we expand the \texttt{BCO Phase Space} approach to include low-mass compact objects (LMCOs), which can be particularly interesting to decipher the mass-scale of transition from neutron star to black holes. This class is particularly intriguing as their formation mechanisms remain poorly constrained, especially within the "lower mass gap" (approximately 2.5 to 5 M$_\odot$ 
\citep{Bailyn:1997xt,2010ApJ...725.1918O,2011ApJ...741..103F,2012ApJ...757...91B}), and potentially encompass a diverse array of astrophysical and primordial origins. Typically, low-mass compact objects include neutron star, primarily formed via core-collapse supernovae \citep{Kalogera:1996ci,Mueller:1996pm,2012ApJ...749...91F,Muhammed:2024zmk,Zuraiq:2023bpw}; however, several alternative channels might significantly populate this mass range. For instance, under extreme astrophysical conditions involving significant fallback accretion, black hole might form with masses transitional between typical neutron stars and conventional stellar-mass black holes \citep{Thompson:2018ycv,2020ApJ...890...51E,vandenHeuvel:2020chh,Thompson:2020nbd,Jayasinghe:2021uqb,Barr:2024wwl,Song:2024tqr}. These objects, potentially filling the low-mass gap, may include black holes originating from neutron star mergers \citep{Barbieri:2019bdq,Tsokaros:2019lnx,Gupta:2019nwj,Yang:2020xyi,Mahapatra:2025agb}, termed remnants of binary neutron star.

Additionally, the hypothesized existence of primordial black hole (PBH) provides a compelling explanation for the origin of low-mass compact objects. PBH may form predominantly through density fluctuations in the early universe, arising from scenarios such as inflationary perturbations \citep{Green:2014faa}, first-order phase transitions \citep{Khlopov:1999ys}, cosmic string collapses \citep{Green:2024bam}, or bubble collisions \citep{Escriva:2022duf}. Detection of these PBH would have profound implications, significantly impacting both astrophysics and cosmology \citep{Sasaki:2016jop,Bird:2016dcv,Clesse:2016vqa}. Furthermore, the low-mass regime may harbor exotic compact objects \citep{Bezares:2024btu} such as boson stars \citep{Macedo:2013jja,Visinelli:2021uve}, or quark stars \citep{Mukhopadhyay:2015xhs}, arising from physics beyond standard astrophysical models. 

In this study, we demonstrate the utility of the Phase Space framework by applying it to seven low-mass GW events from the GWTC-3 catalog 
\citep{KAGRA:2021duu}, complemented by the GW230529 event \citep{LIGOScientific:2024elc} detected during the ongoing fourth observing run by the LIGO \citep{LIGOScientific:2016dsl}-Virgo \citep{VIRGO:2014yos}-KAGRA \citep{KAGRA:2020tym} (LVK) collaboration. Though the \texttt{BCO Phase Space} can be constructed in terms of the observable parameters such as mass, spin, tidal deformability \citep{Flanagan:2007ix,LIGOScientific:2018cki}, electromagnetic counterparts \citep{2012ApJ...746...48M,Troja:2017nqp}, and other characteristic phenomena which provide crucial diagnostics that can differentiate between standard astrophysical and exotic non-astrophysical scenarios, we consider only the individual component masses and spins, as well as luminosity distances, since LVK detector sensitivities currently limit precise measurements of other potentially informative parameters such as eccentricity and kick velocities. 
We particularly investigate the formation channels of LMCOs in four distinct scenarios: neutron star (NS), remnants of binary NS, primordial black holes, and low mass astrophysical black hole (ABH) formed under extreme astrophysical conditions. We classify these low-mass GW events based on their joint mass-spin posterior distributions. For each event, we calculate a probability indicating whether it is an NS, remnant of binary NS, low mass ABH, or a PBH. The resulting classification, visually represented in Figure~\ref{fig:ProbEvents}, shows the main result from our paper on the probability of association for each component of the analyzed binary events with different formation scenario. However, as observed, these classifications do not unequivocally favor a single astrophysical channel for most events, due to overlap in the phase space of different formation scenarios within the parameter range accessible to the current dataset. Instead, multiple formation pathways remain plausible, underscoring the inherent degeneracies due to current detector sensitivity and limited event statistics. Furthermore, while we adopt component masses and spins here, the LMCO phase space methodology can be equally effective when applied using alternative GW observables, such as chirp mass and effective spin, as demonstrated in our previous work \citep{Afroz:2024fzp}. We explain this results in details in section~\ref{sec:result} of the paper. 

The paper is organized as follows: In section~\ref{sec:PhaseSpacePar}, we explore the properties of low-mass compact objects originating from different formation channels within the phase space and reconstruct the phase space associated with these formation channels and highlight their distinctive features. Section~\ref{sec:Framework} presents a detailed framework outlining our phase space methodology. In section~\ref{sec:result}, we apply our approach to identify observational support for different formation channels using events from the GWTC-3 catalog. Finally, section~\ref{sec:Conclusion} summarizes our key findings and discusses potential future research directions.

\section{Characterizing Formation Pathways for different scenarios in Phase Space}
\label{sec:PhaseSpacePar}

\subsection{Neutron Star}

NS represent one of the most exotic and extreme forms of matter in the universe, offering unique laboratories for fundamental physics. Understanding their properties such as mass, spin, tidal deformability, and electromagnetic counterparts is essential for probing nuclear physics at ultra-high densities, gravitational physics under extreme conditions, and stellar evolution processes. In particular, many of these properties, including the maximum mass and the tidal deformability, are strongly determined by the equation of state of dense nuclear matter \citep{Hebeler:2013nza,Baiotti:2019sew,Ozel:2016oaf}. Among these properties, the tidal deformability and associated electromagnetic counterparts provide critical distinguishing features from black holes. Specifically, neutron stars exhibit significant tidal interactions during merger, manifesting as non-zero tidal deformabilities measurable through GW signals, and often produce electromagnetic counterparts like kilonovae or short gamma-ray bursts.

Neutron stars typically inhabit the lower end of the compact object mass spectrum, generally below $2.5M_\odot$ depending on the equation of state \citep{Markakis:2009mzp,Ozel:2010bz,Hebeler:2013nza,Baiotti:2019sew,GuerraChaves:2019foa,Greif:2020pju,Riley:2021pdl,Miller:2021qha,Ye:2022qoe,Barr:2024wwl}. 

To systematically characterize neutron star populations, their mass distribution is typically modeled using either a Gaussian or a power-law distribution. The Gaussian mass distribution is defined as:

\begin{equation}
p(m) = \frac{1}{\sqrt{2\pi}\sigma}\exp\left[-\frac{(m-\mu)^2}{2\sigma^2}\right],
\end{equation}
where $\mu$ is the mean and $\sigma$ is the standard deviation, with typical parameter choices $\mu \approx 1.33M_\odot$ and $\sigma \approx 0.09M_\odot$  \citep{Kiziltan2013,Ozel2016}. 
Alternatively, power-law distribution can be employed:
\begin{equation}
p(m) \propto m^{\beta}, \quad \text{for } M_{\min}\le m\le M_{\max},
\end{equation}
with typical parameter choices $M_{\min}\approx 1.1M_\odot$, $M_{\max}\approx 2.2M_\odot$, and a slope parameter $\beta\approx0$ \citep{Ozel2016}. These mass distributions effectively capture observed properties from known neutron star populations.

Spin distributions for neutron stars are constrained by observations of Galactic pulsars as well as gravitational-wave data from binary neutron star (BNS) mergers. Galactic pulsar observations typically show neutron stars spinning with relatively low spin magnitudes ($\chi \lesssim 0.05$), reflecting spin-down mechanisms due to magnetic dipole radiation and gravitational-wave emission \citep{Manchester:2004bp,Lorimer:2008se,Ozel:2016oaf,Tauris:2017omb}. On the other hand, GW observations, such as GW170817 also place constraints on neutron star spins, generally indicating low spin magnitudes consistent with Galactic observations. Consequently, neutron star spin distributions are often modeled using uniform distributions within a moderate range, commonly adopted as $\chi \in [0.0, 0.5]$ to capture the theoretical uncertainty and observational constraints \citep{East:2019lbk,LIGOScientific:2018hze, Landry:2020vaw}. These modeling choices reflect our understanding of neutron star spin evolution influenced by formation channels, magnetic field dynamics, and angular momentum transfer processes in binary systems.

The neutron star merger rate as a function of redshift can be modeled by the delay time distribution defined as the interval between progenitor star formation and eventual merger \citep{Vigna-Gomez:2018dza,Safarzadeh:2019znp,Broekgaarden:2021efa,Zevin:2022dbo,Sgalletta:2023erk,Lehoucq:2023zlt}. Typically, a power-law delay time distribution is employed \citep{Dominik:2014yma,Mukherjee:2021rtw,Karathanasis:2022rtr, Karathanasis:2022hrb}:
\begin{equation}
    p_t(t_d|t_d^{\min},t_d^{\max},\gamma) \propto 
    \begin{cases}
    t_d^{-\gamma} & \text{for }  t_d^{\min} < t_d < t_d^{\max}, \\
    0 & \text{otherwise}.
    \end{cases}
\end{equation}
with typical slopes $\gamma \sim 1$. However, it can vary from $\gamma \sim 1$ and its observational validation is yet to be made. Integrating this delay time distribution with the star formation rate (SFR) history yields the redshift-dependent merger rate:
\begin{equation}
R(z) = R_0 \frac{\displaystyle\int_z^{\infty} p_t(t_d|t_d^{\min},t_d^{\max},\gamma), R_{\mathrm{SFR}}(z_f) \frac{dt}{dz_f} dz_f}{\displaystyle\int_0^{\infty} p_t(t_d|t_d^{\min},t_d^{\max},\gamma), R_{\mathrm{SFR}}(z_f) \frac{dt}{dz_f} dz_f},
\label{eq:DelayMergRate}
\end{equation}
where $R_0$ is the local merger rate density, $z_f$ denotes the redshift at stellar binary formation, and $t$ is the look-back time. The upper limit of integration is formally $\infty$, though the contribution effectively vanishes at high redshift because $R_{\mathrm{SFR}}(z)$ rapidly goes to zero, consistent with the Madau--Dickinson form \citep{Madau:2014bja}.

While the power-law delay time distribution is widely used for its simplicity and ability to capture key trends, it has limitations. It cannot fully account for variations in merger rate evolution due to changes in star formation history over cosmic time. Additionally, assuming a single power-law slope across all redshifts may oversimplify the complexities of stellar and binary evolution, especially in rapidly changing environments \citep{Stegmann:2022ruy,Chattopadhyay:2023pil,Antonini:2024het}. Despite these challenges, this approach captures the key features of neutron star mergers without introducing additional uncertainties. For our modeling, we explicitly adopt a delay-time distribution slope of $\gamma=1$ and a minimum delay time of $500\,\mathrm{Myrs}$, consistent with values commonly used in the literature \citep{Dominik:2012kk, Vigna-Gomez:2018dza}. The resulting merger rate is shown in Figure \ref{fig:NSMergRate}. Due to the limited redshift range up to z $\sim 1$ accessible from the LVK detectors to detect neutron star events, these choices of parameters for the delay time distribution will make the merger rate a power-law.

Metallicity influences the formation of neutron stars by affecting the evolution of their progenitor stars, the core-collapse mechanism, and the resulting mass distribution. In general, lower metallicity environments lead to reduced stellar wind mass loss, allowing progenitors to retain more mass and potentially form slightly heavier neutron stars. Conversely, higher metallicity environments promote stronger winds and result in lighter remnants. However, compared to black hole formation which is highly sensitive to metallicity due to its stronger effect on stellar mass loss and fallback the impact of metallicity on neutron star formation is relatively modest \citep{Giacobbo:2018etu,Santoliquido:2020axb,Gallegos-Garcia:2022lfk}. However, the eight GW events analyzed in this study reside predominantly at low redshifts ($z \lesssim 0.2$), minimizing variations in metallicity and therefore its impact on our mass distribution modeling.

While tidal deformability plays crucial roles in fully characterizing neutron star mergers, we do not explicitly utilize these properties here due to the limited sensitivity of current GW detectors to measure them accurately for most events. Instead, we primarily focus on component masses, spins, and luminosity distances, providing a robust and directly observable foundation for exploring neutron star formation channels in the proposed LMCO Phase Space framework.

\subsection{Remnants of binary Neutron Star}

Binary neutron star mergers are natural candidates for populating the low-mass gap (approximately $2.5-5 M_\odot$) with black hole remnants. However, the formation of these low-mass black holes typically requires environments rich in compact objects and characterized by frequent dynamical interactions, such as globular clusters (GCs) \citep{Rodriguez:2015oxa,Hong:2018bqs}, young stellar clusters \citep{OLeary:2008myb}, nuclear star clusters, and active galactic nuclei (AGN) disks \citep{Tagawa:2019osr, Wang:2021clu,2024arXiv240216948F}.

Globular clusters, in particular, represent promising sites for such dynamical interactions, though the exact contribution of these clusters to neutron star merger rates remains an area of active investigation \citep{Zevin:2019obe,Kremer:2021zrj,Ye:2024wqj}. Initial studies suggested that mass segregation in globular clusters preferentially favored heavier black holes, thereby reducing neutron star merger rates \citep{Sigurdsson:1994ju,Bae:2013fna,Belczynski:2017mqx}. More recent analyses, however, have revealed that massive, core-collapsed globular clusters can significantly increase neutron star merger rates through mechanisms such as direct stellar collisions and tidal captures. Observational support for these dynamical scenarios is found in the characteristics of short gamma-ray bursts and observed Galactic binary neutron star systems, which frequently exhibit eccentric orbits indicative of dynamical origins \citep{Grindlay:2005ym,Guetta:2008qw,2010ApJ...720..953L,Fragione:2018jxd,Ye:2019luh}. Despite these findings, uncertainties persist regarding the precise contribution of various dense stellar environments, though their potential to populate the low-mass gap remains compelling \citep{Sedda:2020wzl,Ye:2019xvf}.

\begin{figure}
\includegraphics[width=8.5cm, height=5.5cm]{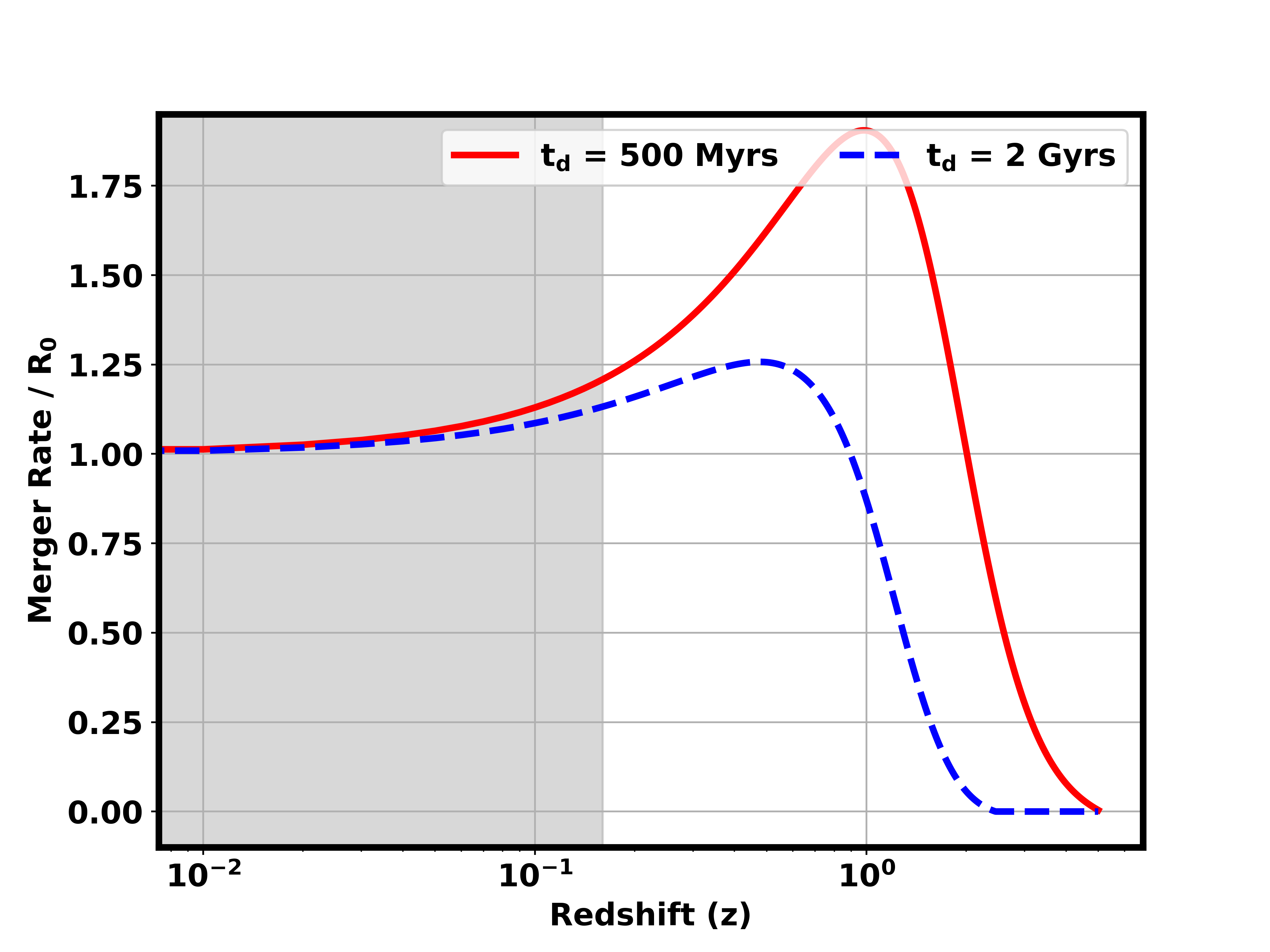}
\caption{Evolution of the normalized merger rate density ($R(z) / R_0$) as a function of redshift ($z$) for two different delay times: $t_d = 500$ Myrs ((solid red line, corresponding to NS) and $t_d = 2$ Gyrs ((dashed blue line, corresponding to NS remnants)). The merger rate peaks at lower redshift for the longer delay time, reflecting the delayed merging of binary black holes formed at higher redshifts. The difference in the shape of the curves highlights the impact of the delay time on the merger rate evolution. The shaded region indicates the redshift range accessible to the LVK detectors for the low-mass binary black hole events considered in this analysis.}
\label{fig:NSMergRate}
\end{figure}

Remnants of BNS occupy a distinct and identifiable region at the lower end of the mass spectrum. Typically, progenitor neutron stars are modeled using Gaussian or flat mass distributions below approximately $2.5M_\odot$. During mergers, gravitational-wave emission leads to the retention of roughly 95\%  of the combined mass of the progenitors \citep{Pretorius:2005gq,Ossokine:2017dge}. 

For our analysis, motivated by observed neutron‐star mass distributions, we adopt a uniform prior on remnant mass in the range \(M\in[2,5]\,M_\odot\).  {Importantly, the phase‐space formalism is agnostic to this choice: substituting any alternative mass model (e.g.\ Gaussian, broken power law, or multimodal distribution) simply reshapes the joint prior \(P_{\rm remnant}(M,\chi)\) and its marginalized projection (for more details, refer to Section \ref{sec:Framework}). Since different mass\text{-}spin combination occupy distinct regions of the \((M,\chi)\) plane, our classification will unambiguously reflect these differences and thus clearly distinguish between competing remnant prescriptions without altering the core framework.}

Remnants of BNS generally exhibit higher spin values due to the significant angular momentum transferred during the coalescence process. In a binary neutron star merger, a portion of the orbital angular momentum is converted into the spin of the remnant object. The highly dynamic and violent nature of the merger characterized by tidal deformation, mass ejection, and differential rotation contributes to the formation of a rapidly spinning remnant. Hydrodynamical simulations have shown that, depending on the equation of state and the mass ratio of the binary, the remnant can exhibit a wide range of spin characteristics \citep{Bernuzzi:2016pie, Shibata:2019wef, Radice:2020ddv}. These studies indicate that if the remnant survives for at least a few milliseconds, it can attain a very high dimensionless spin parameter, especially in the absence of prompt collapse. Strong differential rotation and the redistribution of angular momentum further amplify the spin \citep{East:2015vix,Kastaun:2016elu,Dietrich:2016lyp,Karakas:2024avr}. This contrasts with neutron stars or black holes formed through isolated stellar evolution, which typically have lower spins due to angular momentum loss through stellar winds or magnetic braking. Additionally, these mergers often take place within dynamically active environments, where interactions with surrounding matter and gravitational effects further contribute to higher spin magnitudes \citep{Campanelli:2006uy,Scheel:2008rj,Fishbach:2017dwv,Ye:2024wqj,Xing:2024ydg}.

The redshift distribution of remnant of BNS depends sensitively on their host environments. Variations in the merger rate arise from the distinct dynamical histories and evolutionary timescales in different stellar environments. Typically, the merger rate within clusters can be approximated by scaling the global star formation rate (SFR) density, adjusted by an exponential decay factor that accounts for cluster disruption and evaporation over cosmic time. However, given the large theoretical uncertainties in modeling these environmental dependencies and the fact that all eight GW events analyzed in this study occur at low redshifts (\(z \lesssim 0.2\)) these effects are significantly reduced in our analysis \citep{Mennekens:2016jcs,Beniamini:2019iop,Chu:2021evh}. In the low-redshift regime, the overall shape of the neutron star remnants merger rate resembles a simple power-law, though its precise form remains unconstrained due to the limited redshift range of currently observed events. Therefore, to broadly account for environmental and evolutionary delays, we adopt a fixed delay time of 2 Gyrs in our modeling of neutron star merger remnants for this analysis, which can be easily varied in the phase space framework. The resulting merger rate is shown in Figure~\ref{fig:NSMergRate}. Here we use the same progenitor distributions for NS and NS remnants, and vary only the delay time to illustrate its impact on the merger rate evolution.

In this context, while delay times depend on cluster parameters such as mass, radius, and escape velocity \citep{Stegmann:2022ruy,Chattopadhyay:2023pil,Antonini:2024het}, our approach captures the key features of the mergers without introducing additional uncertainties from detailed cluster modeling. Moreover, while this model is sufficiently general to represent various merger scenarios, including those influenced by cluster dynamics, the current dataset comprising only a few events cannot yet fully constrain all its detailed dependencies. Nor can it constrain more complex merger rate models incorporating additional astrophysical details. Nevertheless, for any alternative merger rate form, a phase space can be constructed accordingly.

\subsection{Primordial Black Holes}

Primordial black holes are potential sources of GW that could be detected by current and future GW observatories \citep{Sasaki:2016jop,Bird:2016dcv,Clesse:2016vqa}. Their merger rate, mapped from redshift to luminosity distance ($D_L$), is highly sensitive to their formation scenarios. Depending on different formation mechanisms \citep{Crawford:1982yz,Garcia-Bellido:1996mdl,Jedamzik:1996mr,Musco:2012au,Raidal:2024bmm}, the growth of a PBH’s mass is significantly influenced by its initial mass distribution. Over time, PBH can further increase their mass through matter accretion \citep{Mack:2006gz,Serpico:2020ehh}. The maximum rate at which a PBH can accrete matter is defined by the Eddington limit, balancing gravitational attraction and radiation pressure. This accretion rate is expressed as \citep{Franciolini:2021nvv}:
\begin{equation}
\dot{M} = \dot{m}\,\dot{M}_{\mathrm{Edd}} = \dot{m} \times 2.2 \, \frac{M_{\odot}}{\text{Gyrs}} \left(\frac{M}{M_{\odot}}\right) \,,
\label{eq:MassEvoPBH}
\end{equation}
where $\dot{M}_{\mathrm{Edd}}$ is the Eddington accretion rate, $M$ denotes the mass of the PBH, and $\dot{m}$ is the dimensionless mass accretion index. For PBH with very low initial masses (of order $10^{-5}\,M_\odot$) and high accretion rates, the final masses can grow to fall within the sensitivity range of current GW detectors over the age of the Universe. Conversely, PBH starting with higher masses (around $1\,M_\odot$) but with lower accretion rates can also evolve to detectable masses.

A value of $\dot{m} = 1$ corresponds to Eddington-limited accretion, whereas values below 1 indicate sub-Eddington rates. Moreover, there exists a relationship between mass accretion and spin evolution. As PBH accrete matter, the angular momentum transferred by the infalling material increases their spin. The spin parameter, $\chi$, evolves from an initial value of zero toward a maximum of one, depending on both the redshift $z$ and the accretion rate $\dot{m}$. We parametrize this evolution as
\begin{equation}
\chi(z, \dot{m}) =  1 - e^{-k\,\dot{m}\,\Delta t} \,,
\label{eq:SpinEvoPBH}
\end{equation}
where $\Delta t$ is the time interval (in Gyrs) between an initial redshift $z_{\text{initial}}$ and the current redshift $z$, and $k$ is a constant controlling the rate of spin growth. Different values of $k$ yield varying spin evolution rates.

The merger rate of PBH is given by \citep{Raidal:2018bbj, Vaskonen:2019jpv}:
\begin{equation}
\begin{aligned}
\mathrm{dR_0} = \frac{1.6\times10^6}{\mathrm{Gpc^3\,yr}}\,f_{\mathrm{PBH}}^{\frac{53}{37}}\eta^{-\frac{34}{37}}\left(\frac{M}{ M_\odot}\right)^{-\frac{32}{37}}\left(\frac{\tau}{t_0}\right)^{-\frac{34}{37}} \\
\times \, 0.24\left(1+\frac{2.3S_{eq}}{f_{\mathrm{PBH}}^2}\right)\psi(m_1)\psi(m_2)\,dm_1\,dm_2 \,,
\end{aligned}
\end{equation}
where $\tau$ is the time elapsed since PBH formation, $t_0$ is the current age of the Universe, and the factor $\left(1+\frac{2.3S_{eq}}{f_{\mathrm{PBH}}^2}\right)$ accounts for the influence of adiabatic perturbations on the eccentricity of early binaries, with $\sqrt{S_{eq}} = 0.0005$ representing the variance of matter perturbations at matter–radiation equality. Here, $f_{\mathrm{PBH}}$ denotes the fraction of dark matter comprised of PBH, $\eta$ is the symmetric mass ratio, and $M$ is the total mass of the binary. The mass distribution $\psi(M)$ is central to understanding the PBH population and is normalized so that its integral over all masses yields the total PBH density.

A widely used approximation for the PBH mass distribution is the log-normal distribution \citep{Carr:2018rid}:
\begin{equation}
\psi(M) = \frac{1}{\sqrt{2\pi}\,\sigma\,M} \exp\left[-\frac{\ln^2(M/M_c)}{2\sigma^2}\right] \,,
\end{equation}
where $M_c$ is the characteristic mass and $\sigma$ defines the width of the distribution. This parameterization can be readily adjusted to meet observational constraints and explore diverse PBH populations \citep{PhysRevD.47.4244,Green:2016xgy,Kannike:2017bxn,Kuhnel:2017pwq}.

While we adopt the log-normal distribution as it is widely used in the literature and provides a simple two-parameter characterization of PBH masses, alternative thermal-history dependent mass functions have also been proposed \citep{Carr:2019kxo}. Such distributions incorporate additional early-Universe physics and may lead to different predictions for the PBH population.

In this work we restrict our analysis to the early-Universe PBH binary formation channel. Late-Universe dynamical formation channels are not explicitly modeled here, as they are generally expected to contribute subdominantly in most PBH scenarios and would require additional astrophysical modeling. Moreover, given the very limited number of GW events analyzed in this study, including such channels would not significantly change our conclusions. Our phase-space approach could in principle be extended to disentangle the two channels, since their redshift distributions are expected to differ \citep{Raidal:2018bbj,Ali-Haimoud:2017rtz,Franciolini:2022ewd, Raidal:2024bmm}.

\subsection{Low-Mass Astrophysical Black Holes}

In standard astrophysical scenarios, black holes typically form with masses above a lower mass cutoff, below which it can be a neutron star. This mass-scale is quite uncertain due to ignorance about the astrophysical processes and also the equation of state of neutron stars. Typically a mass approximately \(5\,M_\odot\) is assumed for black holes for GW analysis. However, under extreme conditions such as inefficient shock revival, significant fallback accretion, or suppressed neutrino heating the outcome of core collapse can differ substantially \citep{Fryer:1999mi,Janka:2013hfa,Chan:2020lnd}. In these scenarios, a large fraction of stellar material may fall back onto the proto-compact remnant, resulting in black holes with masses significantly lower than the canonical threshold. This mechanism can populate the so-called low-mass gap, producing black holes within roughly $2 -5 M_\odot$ \citep{Schroder:2018hxk,Tang:2021bnp,Olejak:2022zee}.

One can model the final black hole mass, \(M_{\mathrm{BH}}\), as a function of the pre-supernova stellar mass, \(M_{\mathrm{pre}}\), and the fallback fraction, \(f_{\mathrm{b}}\), as follows:
\begin{equation}
    M_{\mathrm{BH}} = M_{\mathrm{proto}} + f_{\mathrm{b}}\,(M_{\mathrm{pre}} - M_{\mathrm{proto}})\,,
\end{equation}
where \(M_{\mathrm{proto}}\) represents the proto-compact remnant's mass immediately after core collapse, and \(f_{\mathrm{b}}\) (with \(0 \leq f_{\mathrm{b}} \leq 1\)) quantifies the fraction of the stellar envelope accreted onto the remnant. Higher values of \(f_{\mathrm{b}}\) under extreme supernova conditions lead to remnant masses in the low-mass gap \citep{Belczynski:2016jno,Pan:2020idl,Gao:2022fsg,Ghodla:2023ymi}.

To capture the diversity of possible mass outcomes, we parameterize the mass distribution as a combination of a Gaussian core and an exponential tail. Specifically, the median and width of the Gaussian component depend explicitly on the fallback fraction, \(f_{\mathrm{b}}\), reflecting the efficiency of envelope ejection during core collapse. The mass function, \(P(m| f_{\mathrm{b}})\), is thus defined as:

\begin{equation}
P(m|f_{b}) = 
    \begin{cases}
        \frac{1}{\sqrt{2\pi \sigma(f_{b})^2}} \exp\left(-\frac{(m - M_{\rm median}(f_{b}))^2}{2 \sigma(f_{b})^2}\right), &  \\
        \qquad \text{if } m < M_{\rm median}(f_{b}), \\[10pt]
        \frac{1}{\sqrt{2\pi \sigma(f_{b})^2}} \exp\left(-\frac{(m - M_{\rm median}(f_{b})^2}{2 \sigma^2}\right) & \\
        \qquad \times \exp(-\alpha (m - M_{\rm median}(f_{b}))), & \\
        \qquad \text{if } m \geq M_{\rm median}(f_{b}).
    \end{cases}
    \label{eq:MassModel}
\end{equation}

where the median mass, \(M_{\mathrm{median}}(f_{\mathrm{b}})\), and the width, \(\sigma(f_{\mathrm{b}})\), are parameterized as:
\begin{align}
    M_{\mathrm{median}}(f_{\mathrm{b}}) &= M_0\,[1 - \beta\, f_{\mathrm{b}}]\,, \\
    \sigma(f_{\mathrm{b}}) &= \sigma_0\,[1 + \beta\, f_{\mathrm{b}}]\,.
\end{align}
Here, \(M_0\) is the canonical minimum black hole mass, \(\sigma_0\) is the nominal width, and the dimensionless parameters \(\beta\) quantify how the median mass and width of the distribution respond to increased fallback fractions. The parameter \(\alpha\) controls the steepness of the exponential tail at higher masses. This flexible model enables exploration of how variations in \(f_{\mathrm{b}}\) influence the formation of low-mass black holes, and fitting it to observational data provides insights into supernova physics and fallback processes.

The spins of these low-mass astrophysical black holes are significantly influenced by fallback dynamics. While initial spins arise from the angular momentum of the collapsing stellar core, they can be further enhanced by the angular momentum transferred from fallback material \citep{Muller:2023byy,Sykes:2024jwv}. Although the final spins under extreme fallback conditions tend to remain moderate, their precise values depend sensitively on the angular momentum distribution of the accreted matter. In our analysis, we adopt a uniform distribution for the dimensionless spin parameter, \(\chi\), ranging from 0 to 0.5.


The redshift distribution of these low-mass astrophysical black holes, defined by the merger rate, follows the Madau\text{-}Dickinson star formation rate integrated with a delay-time distribution as described by Equation~\eqref{eq:DelayMergRate}. While metallicity typically affects stellar-mass black hole formation by altering mass loss rates (with lower metallicity environments favoring higher masses), the eight GW events analyzed in this study reside at low redshifts (\(z \lesssim 0.2\)), thus mitigating significant metallicity impacts. In modeling, we adopt a standard delay-time distribution slope of \(d=1\) and a minimum delay time of \(500\,\mathrm{Myrs}\), consistent with common astrophysical assumptions. Given that all the low-mass sources considered are observed at low redshifts, our assumption regarding the delay-time distribution will not substantially impact the final results, as the redshift dependence is not strong in this regime.

\subsection{Model Summary}

These four formation channels exhibit distinct features in the \((M,\chi,D_L)\) phase space (along with the other observable features such as tidal deformation), with the following key discriminants\footnote{The range of the parameters mentioned below are underlying prior choices and can be chosen much broader in the phase space analysis}:

\begin{itemize}
  \item \textbf{Neutron Stars:} Masses concentrated at the very low end of the compact‐object spectrum ($1\text{-}2.5M_\odot$), together with minimal spins. Detectable across a broad range of luminosity distances from the local Universe out to moderate redshifts and uniquely identifiable via electromagnetic counterparts and measurable tidal deformabilities. 
  
  \item \textbf{Remnants of Binary Neutron Star:} Remnant of BNS masses lie just above the NS range ($2\text{-}5M_\odot$), with spins enhanced by the merger dynamics. These sources appear at low luminosity distances (low redshifts) due to Gyrs-scale delay times. 
  
  \item \textbf{Low‐Mass Astrophysical Black Holes:} Populate the transitional mass gap with fallback‐driven spin up, yielding a broader, moderate‐spin distribution.  Their redshift distribution follows the cosmic star‐formation history, extending to larger luminosity distances in line with the peak of stellar birth.  
  
  \item \textbf{Primordial Black Holes:} Can span sub‐solar to several‐solar masses depending on the primordial spectrum, exhibiting a mass-spin-distance correlation that is disjoint from all astrophysical channels and typically appears at lower distances given current detector sensitivities.  
\end{itemize}

The above list summarizes the unique region in the phase space using \((M,\chi,D_L)\) which is captured by these different formation scenarios. So, even though the posterior from individual GW event on the source parameters are not sufficient to give an insight on the formation scenarios, their projection on the phase space of the binaries can show which sources overlap with which region of the parameter space across cosmic distances. Due to the underlying physical correlation between the phase space parameters across cosmic distances which originate from different formation scenarios, we can find probabilities (or weights) of possible formation scenarios by integrating over the phase space volume taking into account the selection effect and measurement uncertainties.

\begin{figure*}
\centering
\includegraphics[height=8.0cm, width=10.0cm]{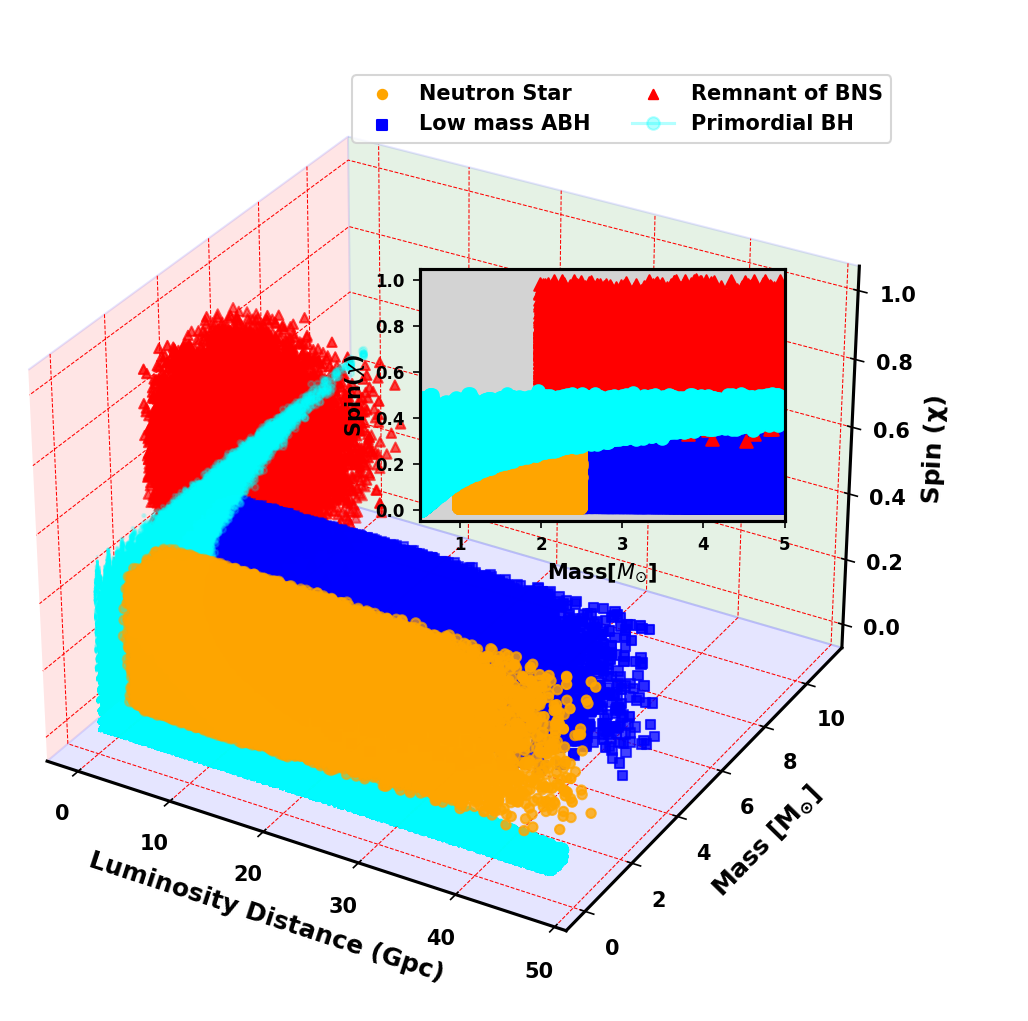}
\caption{This 3D plot illustrates the phase space distribution of Neutron Star, Low mass ABH, Remnant of BNS and PBH, highlighting the distribution of mass, luminosity distance, and dimensionless spin. It shows the distinct regions in the phase space occupied by these compact object. The inset presents the 2D mass-spin projection, further emphasizing the unique mass-spin correlations characteristic of each channel.}
\label{fig:PhaseSpace}
\end{figure*}

\subsection{Reconstruction of Phase Space for Low-Mass Compact Objects}

To systematically reconstruct the phase space for low-mass compact objects, we partition the redshift interval between $z = 0$ and $z = 5$ into 200 evenly spaced bins, each with a width of $\Delta z = 0.025$. We restrict the redshift interval between $z = 0$ and $z = 5$ because contributions from higher redshifts are negligible due to the rapid decline of the star formation rate and the associated delay-time distributions. Moreover, current and near-future GW detector networks are not sensitive to low-mass compact object mergers at higher redshifts.  The total number of GW events in the redshift range $[z_{\rm initial}, z_{\rm final}]$for each compact object population is computed using the integral
\begin{equation}
N_{\mathrm{events}}=T_{\mathrm{obs}} \int_{z_{\rm initial}}^{z_{\rm final}} \frac{R(z)}{1+z}\frac{dV_c}{dz}dz,
\label{eq:TotEvent}
\end{equation}
where $\frac{dV_c}{dz}$ represents the differential comoving volume, $R(z)$ denotes the merger rate specific to each formation scenario, and $T_{\mathrm{obs}}$ is the total observational time period. For each compact object category, we generate events by sampling their mass and spin values from the respective probability distributions using the cumulative distribution function method. Figure \ref{fig:PhaseSpace} is generated solely to illustrate how different compact object formation channels occupy distinct regions of the phase space, and should not be interpreted as a prior for our analysis. The parameters used to produce this phase space are:

\texttt{Neutron Star:} We assume flat distributions for spin and mass, with spins uniformly sampled between 0.0 and 0.4, and masses uniformly distributed between $1$ and $2.2\,M_\odot$. For the redshift distribution, we employ a delay time of $500\,\mathrm{Myrs}$.

\texttt{Remnant of BNS:} The dimensionless spin parameter \(\chi\) is modeled using a Gaussian distribution with a mean value \(\chi_{\rm mean} = 0.5\) and a standard deviation  0.2. Given the current lack of precise theoretical predictions, we treat \(\chi_{\rm mean}\) as a free parameter to account for a wide range of possible spin magnitudes. The mass distribution is assumed to be uniform between \(2\) and \(5\,M_\odot\), while the redshift distribution reflects a longer delay time of \(2\,\mathrm{Gyrs}\). We explore variations in \(\chi_{\rm mean}\) to infer their most likely values in light of the observed GW data, as discussed in Section~\ref{sec:result}.

\texttt{Primordial Black Hole:} We initially considered PBH masses ranging from $0.1$ to $2~M_\odot$ with mass accretion rates ($\dot{m}$) varying between $0$ and $0.1$. These initial conditions were evolved down to the redshifts corresponding to the observed GW events. PBH mass distributions were then obtained by employing a lognormal probability distribution function (Equation~\eqref{eq:MassEvoPBH}). Similarly, the spin distributions were generated using Equation~\ref{eq:SpinEvoPBH}, adopting a shape parameter $k=10$. The total number of events is then calculated assuming a PBH fraction of \(f_{\mathrm{PBH}} = 0.001\). 

\texttt{Low mass ABH:} The mass distribution follows Equation \ref{eq:MassModel}, with parameters set to a fallback fraction $f_{\mathrm{b}} = 0.4$, canonical mass $M_0 = 8\,M_\odot$, distribution width $\sigma_0 = 1.5\,M_\odot$, exponential decay factor $\alpha = 0.15$, and dimensionless shape parameter $\beta = 0.15$. The spins uniformly sampled between 0.0 and 0.4, and for the redshift distribution, we employ a delay time of $500\,\mathrm{Myrs}$. 

Within this LMCO phase space, some regions exhibit overlaps, particularly between low mass astrophysical black holes formed in extreme astrophysical environments and neutron stars. However, distinguishing features like tidal deformability and electromagnetic counterparts help to break such degeneracies. In the case of degeneracy between neutron star merger remnants and other compact object classes, incorporating a fourth observable orbital eccentricity can provide further discriminatory power. Compact binaries evolving in isolation typically exhibit circularized orbits due to mass loss through stellar winds and GW emission \citep{Hurley:2002rf}. Conversely, binaries formed dynamically or via hierarchical mergers often retain significant eccentricities, reflecting their formation environments and dynamical histories \citep{Antonini:2016gqe}. Additionally, PBH occupy distinct trajectories within phase space, differentiated clearly by their unique evolutionary histories and characteristic mass-spin relations driven by early-universe density fluctuations. Though some overlap with astrophysical black holes may exist, PBH' evolutionary trajectories differ fundamentally from those formed through conventional stellar processes, resulting in distinct observational signatures that aid in their identification within the LMCO phase space framework.

\section{Theoretical Framework of the Compact Object Phase Space Technique}
\label{sec:Framework}

The compact object phase space (introduced in our previous work \citep{Afroz:2024fzp}) is a multidimensional representation designed to map the observable properties of compact objects as inferred from GW observations to the theoretical predictions and verify them from the observations to understand the physics behind the formation and evolution of compact objects. We define the phase space as an \(N\)-dimensional manifold, where each dimension corresponds to a specific observable, such as the component masses, dimensionless spins, and luminosity distance. These observables are critical in capturing the formation and evolution of the compact objects, as each formation scenario leaves distinct signatures through these parameters in phase space. The primary objective of the compact object phase space approach is to systematically identify and characterize these formation channels by directly comparing observational data with theoretical model predictions. Importantly, this framework is not limited to validating known models: it also provides a model-agnostic avenue for discovering previously unrecognized or unexpected formation channels. By highlighting statistically significant clustering, outliers, or unexplained regions in the populated phase space, the technique can reveal novel subpopulations or rare pathways that might be missed by traditional model-driven analyses.

In this study, we consider a reduced parameter space defined by three fundamental observables:
\begin{equation}
    \vec{X} = (M, \chi, D_L).
\end{equation}
Nevertheless, the methodology is inherently flexible and can be extended to higher-dimensional representations when more observables are available or when improved detector sensitivity allows the measurement of additional parameters. Each compact object of a GW event is characterized by a posterior probability distribution over the parameter space, denoted as:
\begin{equation}
    P_{\text{GW}}(\vec{X}) = P(M, \chi, D_L).
\end{equation}
This distribution encapsulates the measurement uncertainties and correlations between the observed parameters, accurately reflecting the limitations and precision of the detection process. In this way, the full parameter-estimation uncertainties and correlations from each GW event are carried forward into the phase space representation, ensuring that the reconstructed distributions accurately reflect observational limitations. The total phase space density from a single GW event is then constructed as:
\begin{equation}
    Z(\vec{X}) = \frac{P_{\text{GW}}(\vec{X})}{\int P_{\text{GW}}(\vec{X}) \, d\vec{X}},
\end{equation}
where the normalization ensures that the integrated probability density equals unity:
\begin{equation}
    \int Z(\vec{X}) \, d\vec{X} = 1.
\end{equation}

When multiple GW events are analyzed, each event contributes a separate probability distribution to the phase space. To construct a cumulative phase space that consolidates observational information from all available events, we perform a summation over the individual normalized distributions as 
\begin{equation}
    Z_{\text{total}}(\vec{X}) = \sum_i Z_i(\vec{X}),
\end{equation}
where \(Z_i(\vec{X})\) is the normalized probability distribution associated with the \(i\)-th event. Since the distribution \(Z_i(\vec{X})\) inherently accounts for the measurement uncertainties within the posterior probability density \(P_{\text{GW}}(\vec{X})\), the sum naturally accumulates observational support from multiple events. The aggregation of multiple events in this way reflects the total observational support, making efficient use of the available data while maintaining the relative contributions from each individual event.

In order to connect the observed GW sources with the possible formation channel, we need to project the data on the theoretical trajectories in the phase space. The theoretical trajectories within the compact object phase space represent evolutionary pathways of binary systems driven by distinct formation mechanisms. The idealized trajectory can be expressed as:
\begin{equation}
    T(\vec{X} \mid \{\lambda\}) = P_{\text{model}}(\vec{X} \mid\{\lambda\}),
\end{equation}
where \(P_{\text{model}}(\vec{X} \mid \{\lambda\})\) denotes the probability distribution for the given model parameters denoted by $\{\lambda\}$. In realistic scenarios, the intrinsic scatter in physical processes is inherently encapsulated within the model parameters themselves. Therefore, the uncertainty associated with each trajectory naturally arises from the stochastic nature of the formation mechanism and the distribution of model parameters.

To determine the likelihood of a given formation channel, we project these theoretical trajectories onto the aggregated phase space derived from observations. The probability associated with a specific formation channel is computed as:
\begin{equation}
    \mathrm{P_{\text{channel}}(\{\lambda\})} \propto \int \, Z_{\text{total}}( T(\vec{X} \mid \{\lambda\})) \, d\vec{X}.
\end{equation}
This integral quantifies the weighted overlap between the theoretical trajectory and the observed data distribution, effectively measuring how well the theoretical model explains the observed data. The term \(T(\vec{X} \mid \{\lambda\})\) provides the set of allowed phase space points for that particular formation channel, and \(Z_{\text{total}}( T(\vec{X} \mid \{\lambda\}))\) calculates the cumulative probability from the total observational phase space distribution at those specific points.

Accounting for selection effects is essential in GW analysis because GW detectors have limited sensitivity and are inherently biased toward detecting more massive or nearby sources. To address these limitations, we introduce a selection function that effectively models the detectability of GW events:
\begin{equation}
    S(\vec{X}) = \Theta(\rho(\vec{X}) - \rho_{\text{th}}),
\end{equation}
where \(\rho(\vec{X})\) denotes the signal-to-noise ratio (SNR) as a function of the observed parameters and \(\rho_{\text{th}}\) is the detection threshold. Here, \(\Theta\) is the Heaviside step function defined as:
\begin{equation}
    \Theta(x) = 
    \begin{cases} 
      1 & \text{if } x \geq 0, \\
      0 & \text{if } x < 0.
    \end{cases}
\end{equation}

The selection function effectively filters out regions of the phase space where the GW detectors lack sensitivity, thereby retaining only the physically detectable trajectories. This treatment ensures that undetectable sources do not artificially populate the phase space and bias our inference. By applying the selection function directly through the SNR threshold, we consistently account for the detector sensitivity and restrict the analysis to systems that would be physically observable. This allows us to incorporate selection effects in a way that is both robust and consistent with the current GW detection framework, ensuring that the reconstructed phase space reflects only detectable populations.

Therefore, the effective trajectory likelihood, considering selection effects, becomes:
\begin{equation}
\mathrm{P_{\text{channel}}(\{\lambda\})} \propto \int \, Z_{\text{total}}( T(S(\vec{X}) \mid \{\lambda\})) \, d\vec{X}.
\label{eq:channelweight}
\end{equation}
This formulation guarantees that only physically detectable trajectories contribute to the final probability calculation. It naturally accounts for the detector's sensitivity limitations, ensuring that the inferred formation channels are consistent with the observational capabilities of the GW detectors.

The compact object phase space technique thus provides a powerful and flexible framework to analyze GW events systematically and find the chances of GW sources to arise from different formation channels in a physics-driven way. Its primary strength lies in its ability to 
connect the observations with the theoretical models based on physics-driven model allowing to 
identify promising formation scenarios as well as possible unknown channels (or outliers). This method is based on a Bayesian framework and incorporate uncertainties and observational biases while enabling the inference robust. As more GW detections become available, this approach will continue to be instrumental in unraveling the complex formation pathways of binary compact objects. The key advantages of the method are listed below:

\begin{itemize} 
    \item \textbf{Joint exploration of the observable and theoretical spaces:} The method captures correlations across the full observable parameter space and enables direct mapping to theoretical models, making it possible to study the interplay between measured quantities and model predictions in a unified framework.

    \item \textbf{Physics-driven inference of channel weights:} By comparing model trajectories against the observational phase space, this technique facilitates the estimation of relative contributions from different formation channels in a physically motivated manner. It also naturally incorporates selection effects and detector biases via forward modeling.

    \item \textbf{Discovery potential for new phase space trajectories:} The framework allows for model-agnostic identification of outliers and unexplained clustering in the observed phase space, potentially revealing new or rare formation scenarios not captured by existing models.

    \item \textbf{Computational efficiency and scalability:} Once the data-informed phase space density is constructed, testing different models reduces to efficient integrations obtained by projecting onto different trajectories. This makes it considerably faster and computationally inexpensive. Moreover, the framework is scalable: while the present study focuses on a three-dimensional phase space, the method extends naturally to higher dimensions, with computational cost growing much more mildly than in traditional likelihood-based approaches.
    
\end{itemize}

This makes the phase space approach a robust and forward-looking tool for analyzing the growing catalog of GW events.

\begin{figure*}
    \centering
    \includegraphics[height=7.5cm, width=10.5cm]{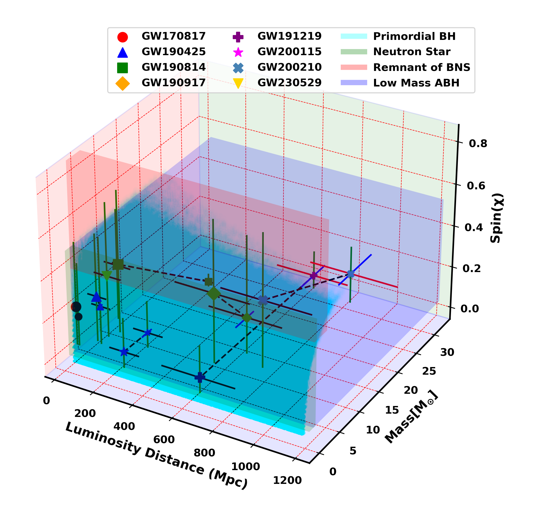}
    \caption{Three-dimensional representation of gravitational-wave events from GWTC-3, along with the newly reported GW230529 from the fourth observing run, for which at least one component’s mean mass lies below \(5\,M_\odot\). The axes display the luminosity distance (in Mpc), component mass (in \(M_\odot\)), and dimensionless spin. Larger circles denote the heavier components, while smaller circles represent the lighter components. Error bars reflect the measurement uncertainties in distance, mass, and spin. Dashed black lines connect the two components of each binary. Shaded regions indicate the parameter spaces corresponding to different formation channels (e.g., primordial black holes, neutron stars, remnants of binary neutron star mergers, and low-mass black holes).}
    \label{fig:catalogmassdist}
\end{figure*}

\begin{figure*}
\centering
\includegraphics[width=0.45\textwidth, height=4.5cm]{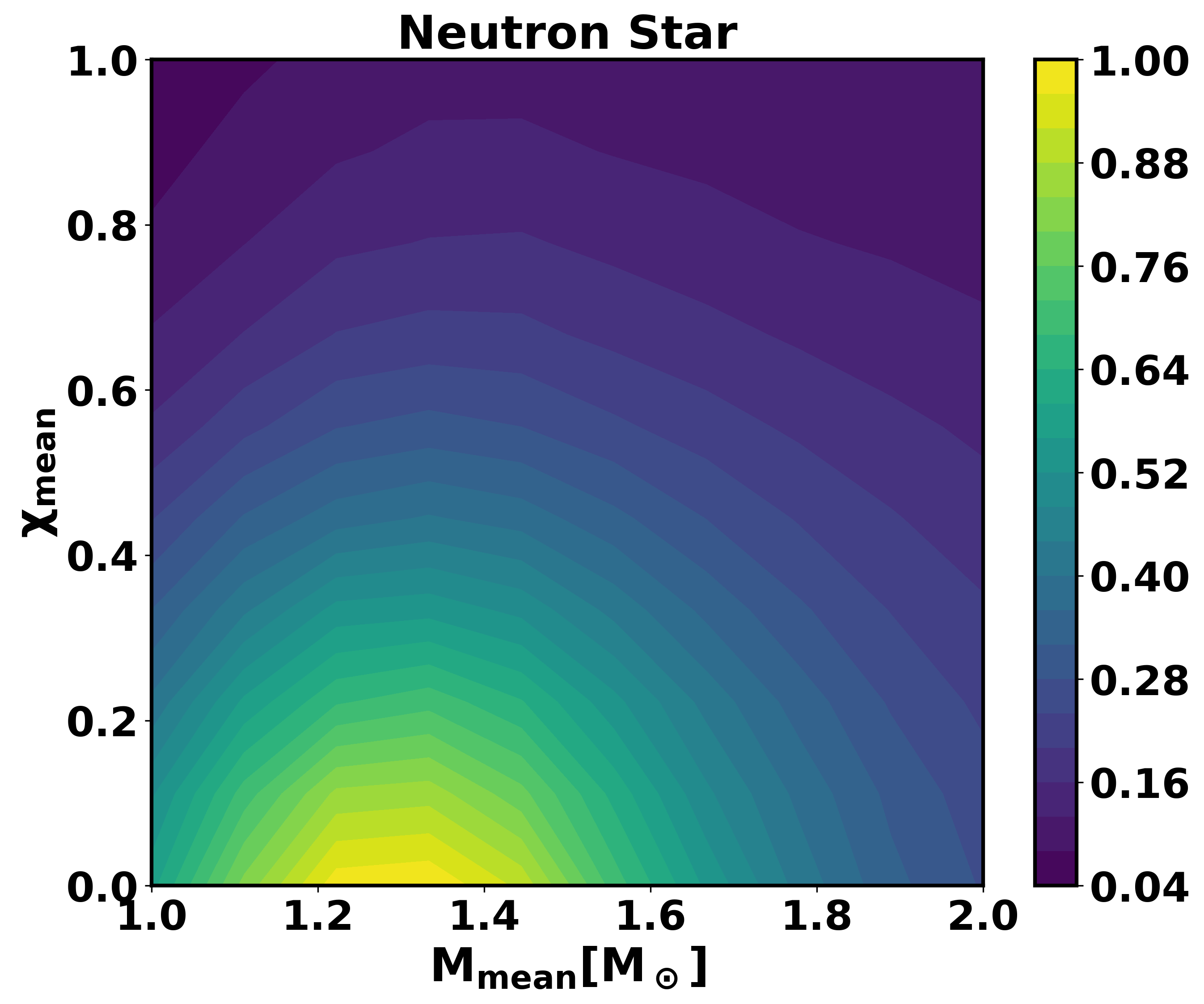}
\includegraphics[width=0.45\textwidth, height=4.5cm]{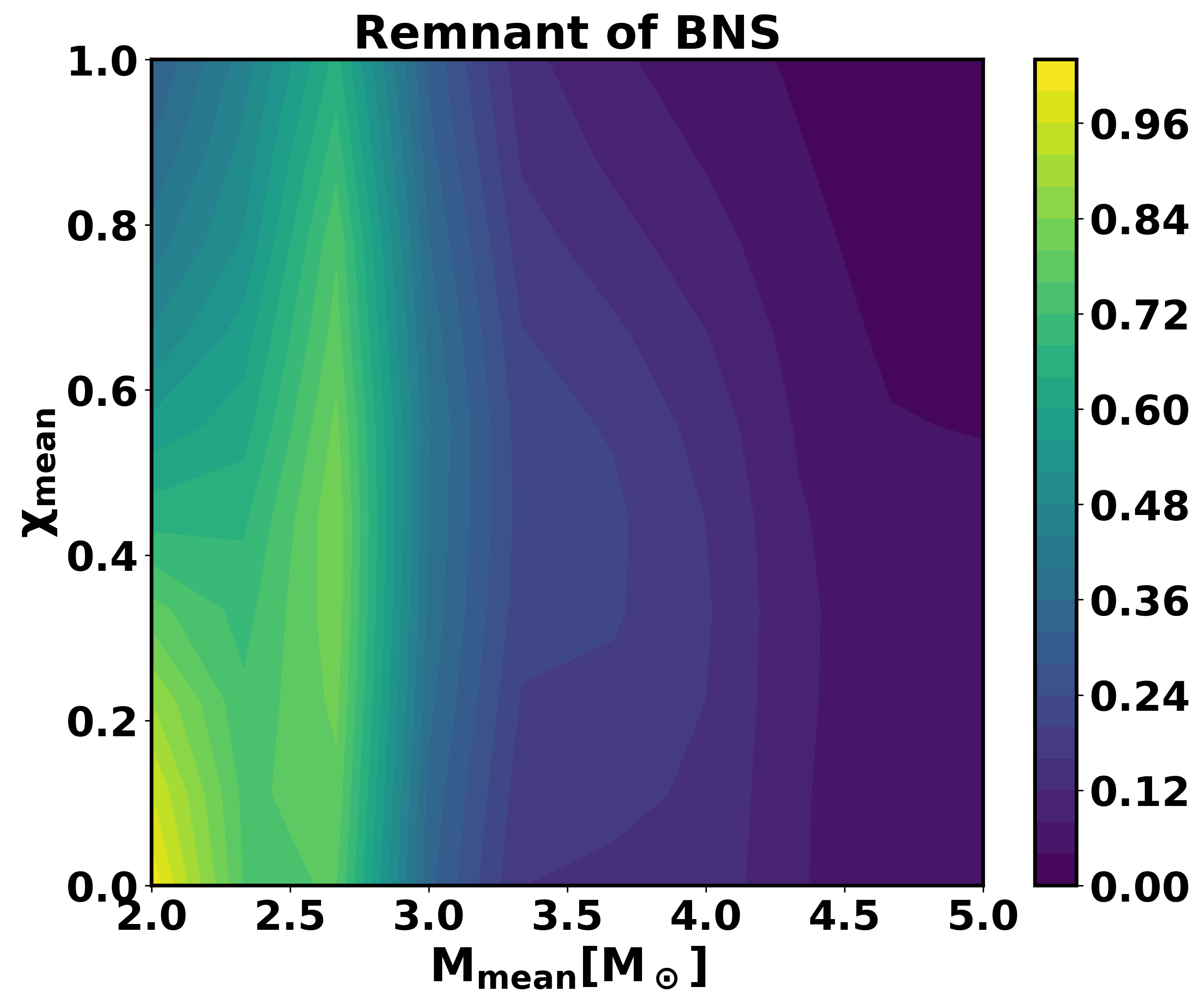}
\includegraphics[width=0.45\textwidth, height=4.5cm]{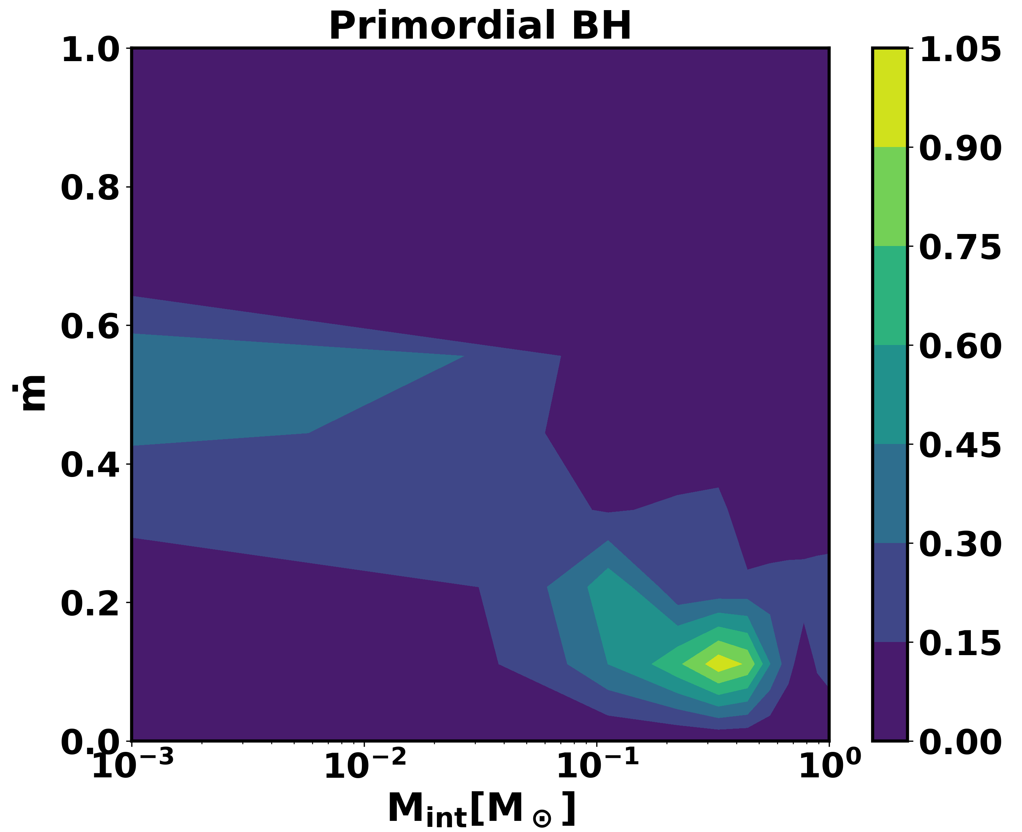}
\includegraphics[width=0.45\textwidth, height=4.5cm]{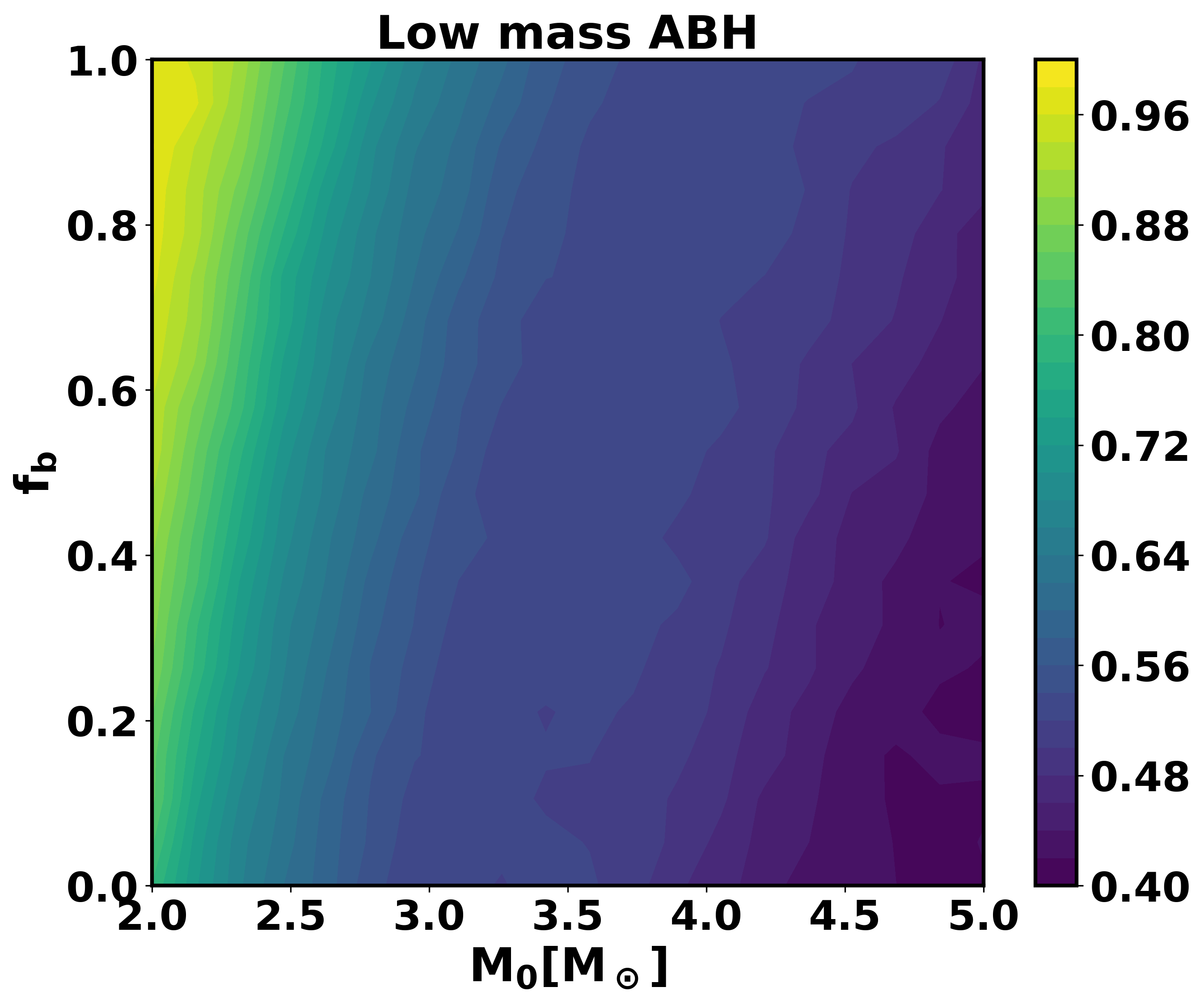}

\caption{
Projection of phase space trajectories for various formation channels of low-mass compact objects. Top Left: Neutron Stars – Illustrates the probability distribution of trajectories as functions of the mean spin ($\chi_{\mathrm{mean}}$) and mean mass ($M_{\mathrm{mean}}$). Top Right: Neutron Star Merger Remnants - Depicts the probability in the same ($M_{\mathrm{mean}}$, $\chi_{\mathrm{mean}}$) space, capturing the merger outcomes. Bottom Left: Primordial Black Holes - Displays the probability distribution in the phase space defined by the mass accretion rate index ($\dot{m}$) and the initial mass ($M_{\mathrm{int}}$). In all panels, the color bar represents the probability values. Bottom Right: Black Holes under Extreme Astrophysical Conditions - Shows trajectory probabilities influenced by the fallback fraction ($f_{\mathrm{b}}$) and the canonical black hole mean mass ($M_0$).
}
\label{fig:ProjectionResult}
\end{figure*}

\section{Identifying Formation Channels of Low-Mass Compact Objects from GWTC-3 catalog}
\label{sec:result}

We investigate the formation channels of LMCOs using GW events identified by the LVK collaboration, focusing on seven selected events from the GWTC-3 catalog (GW170817, GW190425, GW190814, GW190917, GW191219, GW200115, GW200210) \citep{KAGRA:2021duu}, supplemented by event GW230529 \citep{LIGOScientific:2024elc} from the ongoing fourth observing run. Each selected GW event contains at least one binary component with a median mass below $5\,M_\odot$. In contrast, our previous study \citep{Afroz:2024fzp} applied this classification framework across the entire GWTC-3 catalog regardless of component mass with a primary focus on high-mass black-hole binaries. The observational data, comprising component masses ($M_1$, $M_2$), spins ($\chi_1$, $\chi_2$), and luminosity distances ($D_L$), are utilized to construct a comprehensive three-dimensional phase space for comparison with theoretical predictions across various formation scenarios. We show in Figure~\ref{fig:catalogmassdist} the luminosity distance, component masses and dimensionless spin for these selected GW events, along with its uncertainties, overlaid on shaded regions that denote the characteristic parameter‐space domains of the four formation channels: PBH, NS, remnants of BNS, and low mass ABH.

We analyze four distinct formation scenarios for low-mass compact objects: neutron stars, remnant of binary neutron star, primordial black holes, and low mass black hole. Though one can consider several different scenarios in the phase space technique, given the limited amount of data of only eight events from LVK, we restrict to only these scenarios. For each scenario, we systematically explore parameter spaces that capture theoretical uncertainties and astrophysical variations. 

For neutron stars, we vary the mean mass ($M_{\mathrm{mean}}$) from $1.0$ to $2.0\,M_\odot$ and mean spin ($\chi_{\mathrm{mean}}$) from $0.0$ to $1.0$, both with standard deviations of $0.2$. For remnant of binary neutron star, we similarly vary mean mass from $2.0$ to $5.0\,M_\odot$ and mean spin from $0.0$ to $1.0$, again with standard deviations of $0.2$ for both mass and spin. In exploring PBH, we systematically vary initial mass ($M_{\mathrm{int}}$) from $10^{-3}$ to $1\,M_\odot$ and accretion rate indices ($\dot{m}$) from 0 to 1, spanning sub-Eddington to Eddington accretion regimes. For low mass astrophysical black holes, we vary the canonical minimum black hole mass ($M_0$) from $2$ to $5\,M_\odot$ and the fallback fraction ($f_{\mathrm{b}}$) from $0$ to $1$ to examine diverse formation conditions.

Figure~\ref{fig:ProjectionResult} displays the phase space probability projections for each formation channel, clearly distinguishing their characteristic features. These projections are constructed from the model-informed phase-space trajectories described in Section~\ref{sec:PhaseSpacePar}, rather than directly from the posteriors of individual events. Specifically, we first build the observed phase space from all GW events following the procedure outlined in Section~\ref{sec:Framework}. For each formation model, trajectories are then generated by varying its characteristic parameters, and the corresponding weights are computed by integrating these trajectories against the observed phase space. The weighted trajectories, when projected onto two-dimensional subspaces, define the regions shown in Figure~\ref{fig:ProjectionResult}. They therefore represent the expected domains occupied by different populations in the mapped phase space and are not Bayesian priors in the conventional sense. In the limiting case of a single event, its posterior would contribute as one normalized density in the phase space, whose projection would appear as a localized region (reflecting parameter uncertainties) rather than as a single point.

The neutron‐star channel (top‐left panel) peaks prominently at \((M_{\rm mean},\chi_{\rm mean})=(1.3\,M_\odot,0.1)\) with 1\(\sigma\) widths of \(0.2\,M_\odot\) in mass and \(0.2\) in spin, closely matching the low‐mass, low‐spin compact objects in our sample.  In particular, this ridge encompasses both components of GW170817 as well as the secondary components of GW190425, GW191219, GW200115, and GW230529. The remnant of binary neutron star channel (top-right panel) shows a dominant peak near $2.0\,M_\odot$ with similarly low spins, also indicative of low-mass, low-spin objects. Additionally, it features a secondary ridge around $M_{\rm mean} \approx 2.6\,M_\odot$, corresponding to the heavier companions in GW190814, GW190917, and GW200210. A weaker tail at higher masses ($3.0$-$3.5\,M_\odot$) arises primarily from the primary component of GW230529 ($m_1=3.64\,M_\odot$), although spin constraints here remain limited. In the low-mass astrophysical black hole channel (bottom-right panel), the distribution peaks at canonical masses around $M_0 \sim 2.0$-$2.2\,M_\odot$ and high fallback fractions $f_{\rm b}\sim 0.8$-$1.0$. This is driven by the presence of light black hole candidates, specifically the secondary components of GW190814, GW190917, and GW200210 (with masses near $2.79\,M_\odot$). Finally, the primordial black hole channel (bottom-left panel) remains largely unconstrained by existing observations since no detected events have component masses below $1\,M_\odot$. Nonetheless, there is a narrow region of interest around $M_{\rm int}\sim 0.1 M_\odot$ and accretion rates $\dot{m}\sim 0.1$, suggesting the possibility of primordial seeds with minimal spin growth, which is also consistent with our findings in \citep{Afroz:2024fzp}. Figure~\ref{fig:ProjectionResult} illustrates the relative statistical support for different trajectories in phase space, as quantified by the weights in Equation~\ref{eq:channelweight}. Darker regions indicate stronger agreement with the model, while lighter regions are progressively less favored. For example, in the NS panel, the area around $1.2$--$1.4\,M_\odot$ and $\chi_{\rm mean}\approx 0$ receives the highest weight, showing that these values are most strongly supported by the current GW data, though other values are not strictly excluded. We note that Equation~\ref{eq:EventlabelWeight} is used only for computing event-level probabilities (Figure~\ref{fig:ProbEvents}), while Figure~\ref{fig:ProjectionResult} is entirely based on the trajectory weights from Equation~\ref{eq:channelweight}.

As possible from the phase space analysis, we classify these low-mass GW events based on their joint mass-spin posterior distributions. For each event, we assign a probability indicating whether it is a neutron star, remnant of BNS, low mass astrophysical black hole, or a Primordial black hole. For each GW event, we computed the probability weight $w_{\text{channel}}$ for each channel using
\begin{equation}
    w_{\text{channel}} = \frac{\int \int P_{\text{event}}(M, \chi)\,P_{\text{channel}}(M, \chi)\,dM\,d\chi}{\int \int P_{\text{channel}}(M, \chi)\,dM\,d\chi},
\label{eq:EventlabelWeight}
\end{equation}
where $P_{\text{event}}(M, \chi)$ represents the joint posterior distribution obtained from GW parameter estimation and $P_{\text{channel}}(M, \chi)$ represents the theoretical mass-spin distributions for each astrophysical scenario obtained by normalizing the marginalized probability projections shown in Figure \ref{fig:ProjectionResult}. In particular, for the low‐mass astrophysical black‐hole channel we adopt a uniform spin prior \(\chi\in[0,0.5]\) and take the mass distribution directly from the marginalized fallback‐fraction \(f_{\rm b}\) and canonical mass \(M_0\) projections in Figure \ref{fig:ProjectionResult}. 

We note that these probability weights $w_{\rm channel}$ can be interpreted as measures of relative statistical support for each formation channel. In this sense, they are conceptually similar to Bayes factors: higher weights indicate stronger support for a given channel compared to alternatives, although we do not explicitly compute Bayes factors in our framework.

Note that we do not include redshift-dependent weights in this calculation, as the events analyzed lie within a narrow redshift range (\(z \lesssim 0.2\)), limiting the discriminating power of redshift. With next-generation detectors such as Einstein Telescope (ET) \citep{Maggiore:2019uih}  and Cosmic Explorer (CE) \citep{Reitze:2019iox}, which will observe low-mass compact binaries out to high redshift, incorporating redshift evolution will become a powerful discriminator. Different formation channels predict distinct redshift-dependent merger rates, for example, neutron stars track the star-formation history with delay times, while primordial black holes trace the dark matter density and can contribute mergers even at very high redshift. These differences become most pronounced beyond $z\sim2$, where the star-formation rate declines while the PBH contribution remains significant. Including redshift-dependent weights in such a regime would greatly reduce degeneracies between channels.

The resulting classification, shown in Figure~\ref{fig:ProbEvents}, displays the assigned probabilities for each binary component across the four formation channels.  Many events remain highly ambiguous under our low‐mass criteria.  For instance, the primary of GW190425 carries substantial weight in the PBH, NS and low-mass ABH channels, while the primary of GW230529 is split between the remnant of BNS and low mass ABH scenarios.  The secondary components of GW230529, GW200115 and GW190425 all favor both the NS and PBH channels, with additional support for low mass ABH formation.  The secondary of GW191219 similarly straddles the PBH and NS hypotheses.  Meanwhile, the secondaries of GW190814 and GW200210 fall predominantly in the low mass ABH and remnant of BNS regions. The blanck circles indicate that the component’s mass lies outside our defined low-mass window (\(1\text{–}5\,M_\odot\)).  Overall, the broad distribution of weights underscores the persistent degeneracies between formation pathways given current detector sensitivities and the limited number of low-mass events.  We have omitted any degeneracy discussion for GW170817, since its binary‐neutron-star origin is unambiguously established by its EM counterpart \citep{LIGOScientific:2017ync}. In future measurement of the tidal‐deformability from BNS can provide additional insights using the phase space technique.

Future gravitational-wave observatories with enhanced sensitivity, such as ET\citep{Maggiore:2019uih,Branchesi:2023mws, Abac:2025saz}, CE \citep{Reitze:2019iox}, and advanced upgrades to the existing LIGO-Virgo-KAGRA \citep{VIRGO:2014yos,LIGOScientific:2016dsl,KAGRA:2020tym} network, will significantly increase detection rates and improve measurement precision. As more events with better-characterized mass-spin distributions become available, our classification methodology will allow us to disentangle these formation channels more clearly and also discover new formation channels which are not theoretically proposed. Ultimately, this advancement will enable a deeper understanding of compact object populations and their origins in our universe.

\section{Conclusions}
\label{sec:Conclusion}

In this work, we extend our \texttt{BCO Phase Space} framework to the Low-Mass Compact Object phase space to investigate the formation channels of low-mass compact objects using gravitational-wave observations. The low-mass regime ($\lesssim 5M_\odot$) is particularly fascinating as it intersects diverse astrophysical processes and exotic theoretical scenarios, presenting unique challenges and opportunities.

By applying our framework to seven selected events from the GWTC-3 catalog \citep{KAGRA:2021duu}, supplemented by event GW230529 \citep{LIGOScientific:2024elc} from the ongoing fourth observing run, each characterized by at least one binary component mass below $5M_\odot$, we systematically explored four key formation scenarios: neutron stars, remnants of neutron star mergers, primordial black holes, and low mass astrophysical black holes. Our approach quantitatively maps the observed GW events onto theoretical evolutionary trajectories, accounting for observational uncertainties and systematically varying model parameters. This data-driven methodology enables a direct comparison between GW event properties and theoretical predictions, allowing us to assess the relative likelihood of different formation channels.

Our key findings demonstrate that neutron star populations strongly favor masses around $1.3M_\odot$ with low spins ($\chi \sim 0.1$). Remnants of neutron star mergers predominantly exhibit masses around $2.0M_\odot$ with similarly low spins ($\chi_{\mathrm{mean}}\sim 0.1$), along with a secondary peak near $2.6M_\odot$. This distribution shows significant degeneracy with slightly heavier neutron stars, complicating the distinction between these scenarios. Low-mass astrophysical black holes cluster at canonical masses of approximately $2.0$-$2.2M_\odot$ under extreme fallback conditions ($f_{\mathrm{b}}\sim 0.8$--$1.0$). Finally, primordial black holes exhibit limited but notable support in a parameter space characterized by initial masses around $ \sim 0.1 M_\odot$ and modest accretion rates ($\dot{m}\sim 0.1$), suggesting potential relevance for certain low-mass, minimal-spin observations.

Our phase space methodology offers distinct advantages over traditional population studies, fully leveraging the  correlations between  multidimensional parameters of GW observations and their intrinsic uncertainties. It provides a flexible, powerful means to systematically test and differentiate between multiple formation channels simultaneously.

While our present analysis employs simplified parametric representations of astrophysical and primordial formation channels - due primarily to current observational limitations - we emphasize that the LMCO phase space approach is inherently extendable to more sophisticated and physically detailed modeling. Future gravitational-wave detectors \citep{Reitze:2019iox,Maggiore:2019uih,Colpi:2024xhw}, with improved sensitivity and frequency coverage, will greatly enhance our ability to differentiate between nuanced formation mechanisms by potentially capturing additional observable parameters such as eccentricity, tidal deformability, and electromagnetic counterparts. 

In particular, incorporating such observables will help to alleviate the degeneracies we find between certain channels — for example, tidal deformability can clearly distinguish neutron stars from black holes, eccentricity carries information about dynamical versus isolated formation, and electromagnetic counterparts provide unambiguous confirmation for neutron star mergers. Moreover, with next-generation detectors observing to higher redshifts, the redshift evolution of merger rates will offer an additional discriminator between astrophysical and primordial channels.

As gravitational- wave astronomy continues to mature and observational datasets expand, our phase space framework stands as a powerful tool for unraveling the complex evolutionary histories of compact objects. Ultimately, this method not only deepens our understanding of known astrophysical processes but also opens avenues for discovering and characterizing new classes of compact objects, offering profound insights into fundamental physics and stellar evolution.

\section*{Acknowledgments}
The authors express their gratitude to Filippo Santoliquido for reviewing the manuscript and providing useful comments as a part of the LIGO publication policy. This work is part of the \texttt{⟨data|theory⟩ Universe-Lab}, supported by TIFR and the Department of Atomic Energy, Government of India. The authors express gratitude to the system administrator of the computer cluster of \texttt{⟨data|theory⟩ Universe-Lab}. Special thanks to the LIGO-Virgo-KAGRA Scientific Collaboration for providing noise curves. LIGO, funded by the U.S. National Science Foundation (NSF), and Virgo, supported by the French CNRS, Italian INFN, and Dutch Nikhef, along with contributions from Polish and Hungarian institutes. This collaborative effort is backed by the NSF’s LIGO Laboratory, a major facility fully funded by the National Science Foundation. The research leverages data and software from the Gravitational Wave Open Science Center, a service provided by LIGO Laboratory, the LIGO Scientific Collaboration, Virgo Collaboration, and KAGRA. Advanced LIGO's construction and operation receive support from STFC of the UK, Max-Planck Society (MPS), and the State of Niedersachsen/Germany, with additional backing from the Australian Research Council. Virgo, affiliated with the European Gravitational Observatory (EGO), secures funding through contributions from various European institutions. Meanwhile, KAGRA's construction and operation are funded by MEXT, JSPS, NRF, MSIT, AS, and MoST. This material is based upon work supported by NSF’s LIGO Laboratory which is a major facility fully funded by the National Science Foundation. We acknowledge the use of the following packages in this work: Numpy \citep{van2011numpy}, Scipy \citep{jones2001scipy}, Matplotlib \citep{hunter2007matplotlib}, and Astropy \citep{robitaille2013astropy}.

\section*{Data Availability}
The corresponding author will provide the underlying data for this article upon request.

\bibliographystyle{mnras}
\bibliography{references}
\label{lastpage}
\end{document}